# Two-step recording–development approaches in laser processing of materials. Photoinduced gold nanoparticles – carbonization

**Andrey Kudryashov,**[a,b] **Ivan Lukichev,**[a] **Nikita Bityurin** [a,*]

[a]Gaponov-Grekhov Institute of Applied Physics of the Russian Academy of Sciences, 46 Ulyanov Street, 603950 Nizhny Novgorod, Russia

[b]N.I. Lobachevsky State University of Nizhny Novgorod, 23 Gagarin Ave, 603950 Nizhny Novgorod, Russia

**Abstract**. A short review on two-step laser processing of material is presented. The main focus is on the processes which can be called recording–development ones. Here, the first step of laser processing provides an initial pattern on the material surface, which is enhanced or developed at the second step. A new type of the recordingd–development process is considered. The first process is the UV third harmonic of an Nd:YAG laser initiated gold nanoparticle growth in a polystyrene film. A grating of 20-μm period is recorded through the corresponding mask. The lines of the grating are of red color corresponding to the plasmon resonance absorption of gold nanoparticles. The second, development step is carbonization of the matrix just near the gold nanoparticles performed by homogeneous irradiation of the recorded pattern by the powerful radiation of the second harmonic of the same laser. As a result, a black carbonized grating is obtained. The patterns possessing both red and black parts are also presented, demonstrating the opportunity to combine within the same matrix the microstructures of nanocomposites of different kinds.

**Keywords**: two-step laser processing, photoinduced gold nanoparticles, laser carbonization, polymer surface patterning

*bit@appl.sci-nnov.ru

## 1 Introduction

Two-step laser processing is a popular approach to surface patterning. There are different ways to apply this technique. Very often at the first step, powerful pulses provide a crude effect, whereas at the second step more gentle pulses improve the initially created structure [1-11]. In another case, the first and the second irradiation steps are successive different operations aimed to obtain a final complex structure [12-17].

The papers on the two-step laser processing where at least one-step is connected with the Laser Induced Periodic Structures (LIPS) deserve separate consideration [18-24]. In these works LIPS are generated by femtosecond pulses. At the first step of laser irradiation, initial structures are recorded, e.g., by direct nanosecond laser writing. At the second step, the femtosecond pulses are employed for decorating these structures by LIPS. The first-step structures also can be LIPS. The second-step LIPS can interfere with the initial structures. The LIPS created at the first step may sometimes be decorated by other processes, e.g., by laser induced metal nanoparticle deposition from the precursor solution.

There is also a kind of two-step laser holographic data storage recording in photorefractive lithium niobate crystals, where the first step is sensitizing a crystal with blue light from a CW-laser [25].

This paper is devoted to another type of two-step processing that may be called recording–development. Here, laser irradiation at the first step records the initial structure, which is enhanced or developed at the second step. The recording–evelopment processes are quite routine in microlithography [26], where the image in a resist is recorded by e-beam, UV, X-ray, extreme UV, etc. irradiation followed by image development by liquid mediated treatment to achieve the desired pattern. In our case, the development step is also provided by laser irradiation.

This approach is similar, to some extent, to the two-step laser crystallization process where the first step initiates nucleation, whereas the second one provides nuclei growth [27, 28].

The laser ablation of the diamond is also of this kind. Here, the first step results in graphitization of the irradiated spot followed by ablation of the blackened area (second step) [29].

The next example is from the femtosecond optics. When a powerful femtosecond pulse irradiates a transparent dielectric with intensity smaller than approximately $10^{14}$ W/cm$^2$, free electrons are generated due to the multi-photon transition from the ground state to the conduction band. These free electrons are further accelerated by the inverse bremsstrahlung process up to impact ionization, leading to avalanche ionization. When using two-color pulses, one of the fundamental frequency (FF) and the other of the second harmonic (SH), the SH more effectively generates seed electrons, whereas the FF is more effective in the inverse Bremsstrahlung [30]. Here, the SH pulse provides the recording process and the FF pulse the development. This is important, e.g., if the material is irradiated through a layer of near-field lenses (dielectric microspheres) [31]. It was demonstrated that here the size of the ablation craters is determined by the SH, whereas the more powerful FF provides the energy needed for material elimination [32]. It is understood that the SH beam is better focused by the microlenses than the FF radiation.

Below we mention two more examples of such recording–development processes. It is known that UV irradiation of an optically transparent polyvinyl chloride film results in its darkening as a result of photochemically induced polyene formation. Polyenes are sequences of conjugated double C=C bonds providing optical absorption from UV up to visual light. The irradiated area looks like a brown spot. Polyenes are obtained due to the reaction of dehydrochlorination [33]. The same reaction can be activated by heating the material. Here, the recording–development process is possible. First, one can record a polyene image by the UV light using a photochemical process. This may be, e.g., part of a grating (periodic structure) or some bell-like spot. It is a recording process. After that, one can homogeneously irradiate this image by a powerful optical pulse within the polyene absorption band, thereby heating the points of the initially recorded image. The homogeneously irradiated image is heated inhomogeneously because the spatial distribution of the optical absorption is inhomogeneous. The higher absorption at the particular point, the higher the elevated temperature, the higher the temperature and the larger the rate of the thermally activated reaction of polyene growth, i.e. the higher the absorption increment. Thus, there is a positive feedback leading to the enhancement of the contrast of the initially recorded image. Such a recording–development process has been studied both experimentally and theoretically in Refs. [34] and [35].

Another example of such a process is related to microlithography. It is photochemical amplification of the image recorded within resist by shorter-wave radiation (e.g., X-ray) [36]. It may happen that X-ray irradiation of the resist together with chain breaking in a positive resist can also produce color centers. These centers irradiated by UV or even visual photons can amplify the recorded image, multiplying the chain breaking process, thus enhancing the sensitivity of the resist to the X-ray irradiation.

In this paper, we demonstrate one more recording–development process. Laser carbonization of polymer materials is one of the popular fields of laser technology. The carbonized areas demonstrate enhanced electrical conductivity and related properties [37-39]. Recent publications are focused on recording highly luminescent carbonized patterns [40-43].

Nanocomposite materials with polymer matrix and carbon nanoparticles are promising for applications in electronics, energy storage, sensitive sensors, and wearable technology [44-47]. Among polymer materials that are newly involved in laser carbonization activity is a classical polymer such as polystyrene. The problem of carbonization of polystyrene by nanosecond laser pulses is that the carbonization needs a high enough temperature, so ablation starts before carbonization. One of possible ways to realize it was described in [48], where gold nanoparticles were embedded within the polystyrene film. Heating these nanoparticles by the second harmonic of a nanosecond Nd:YAG laser operating within the plasmon resonance resulted in polymer carbonization just near the nanoparticles. The temperature of the nanoparticles during the pulse was as high as 2000 K, whereas the average temperature of the polymer film was smaller than the glass transition point. The carbonized polymer containing carbon nanoparticles looked like a black body and the carbonization products demonstrated tunable photoluminescence.

In paper [49], it was shown that gold nanoparticles can be obtained in polystyrene by a photoinduced process of decomposition of a gold precursor dissolved within the material bulk. The decomposition of the precursor resulted in gold nanoparticle growth. The treated material became red. This means that there is an opportunity to record polymer-gold nanocomposite patterns by either direct writing or laser irradiation of the material through a mask. This initially recorded pattern can be further developed up to the final carbon-polymer one.

## 2      Materials and methods

Polystyrene films with a thickness of 200 μm containing wt. 2.5-5 % of Au precursor $(Ph_3P)Au(n-Bu)$ were produced by the casting technique from a toluene solution. Patterns containing Au NPs were recorded by UV irradiation of the sample at a temperature of about 100 °C in a thermostat. The experimental setup is shown in Fig. 1. The temperature of the sample was controlled using the Optris PI400 thermal imager (Optris GmbH, Germany). UV LED NVSU233A (Nichia, Japan) with a central wavelength of 365 nm and the nanosecond Nd:YAG laser (Lotis TII, Minsk, Belarus) operated at the third harmonic (TH) (355 nm) were used as sources of UV irradiation. After obtaining Au NP containing patterns, the sample was irradiated using the second harmonic of the Nd:YAG laser (532 nm). After that, carbonization occurred in the sample areas containing Au NPs.

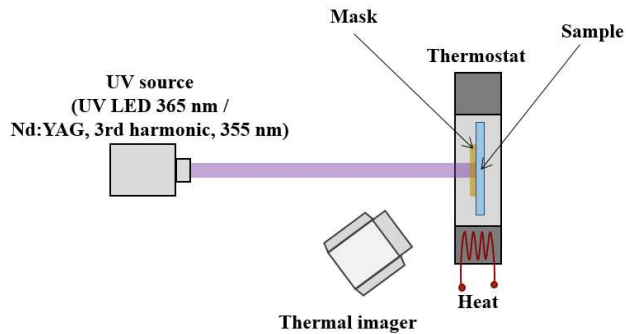 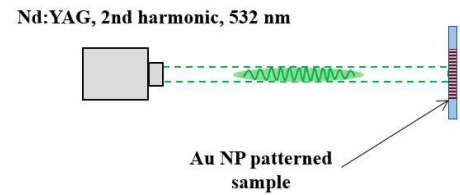

**Fig. 1** Experimental setup for patterning Au NPs and carbonization.

## 3  Results

The first sample (PS + 5% Au precursor) was irradiated with an UV LED (365 nm, 90 mW/cm$^2$) through a diaphragm with a diameter of 1 mm at a temperature of 105°C for 10 minutes. The red spot in Fig. 2(a) is an UV irradiated area, the red color is caused by plasmon resonance in the formed Au NPs. Then, the sample was irradiated with the SH of the Nd:YAG laser (532 nm, 700 mJ/cm$^2$, 40 pulses) focused in the center of the red spot (black spot in the center of the red spot in Fig. 2(a)). The carbonized area exhibits luminescence under UV illumination (Fig. 2(b)).

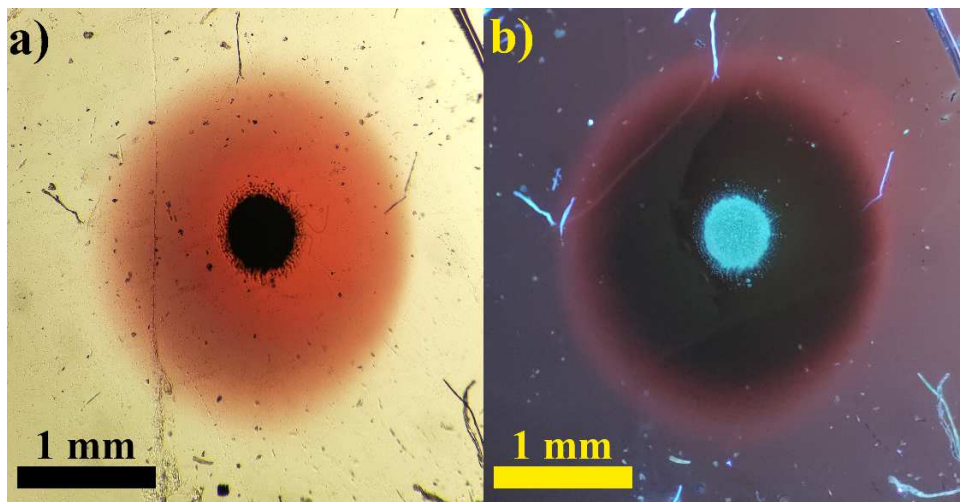

**Fig. 2** Area with Au NPs (red spot) and carbonized area inside it (black spot) in PS film with 5% of Au precursor. A microscopic image when (a) illuminated with white light, (b) luminescence under UV illumination.

The second sample (PS + 2.5% Au precursor) was irradiated with the TH of the Nd:YAG laser (355 nm, 10 Hz, 9 mJ/cm$^2$) at 100°C for 15 minutes. A grating with a period of 20 μm was used as a mask (a microscopic image of the grating is shown in Fig. 3(a)). The result of UV irradiation is presented in Fig. 3(b). The red lines contain Au NPs and have the same period as the grating.

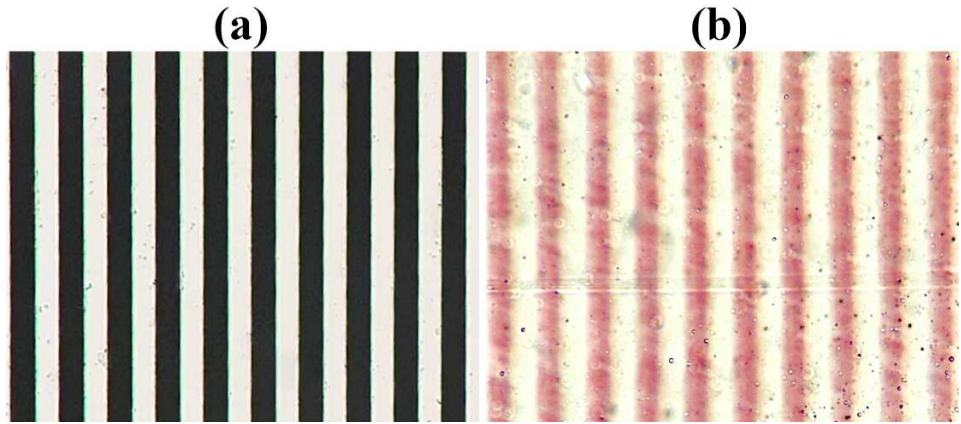

**Fig. 3** A microscopic image of (a) a mask (grating with a period of 20 μm) and (b) a structure with Au NPs recorded through this mask. The period of the recorded structure is the same as of the grating – 20 μm.

Then, the sample with the red lines was irradiated with the SH of the Nd:YAG laser (532 nm, 1 mJ/cm$^2$, 5 pulses). The size of the laser beam was much larger than the thickness of the red lines. Only red lines were carbonized after exposure (fig. 4).

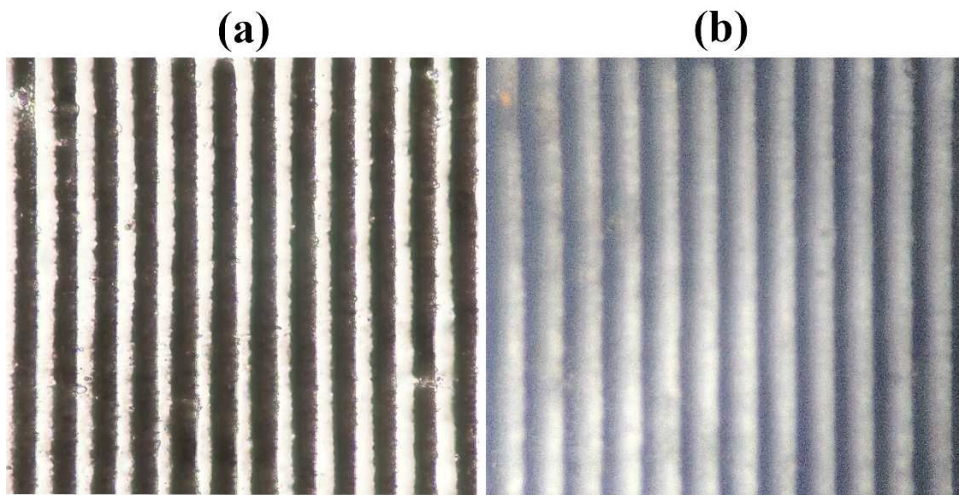

**Fig. 4** Microscopic image of carbonized lines (a) when illuminated with white light, (b) luminescence under UV illumination. The period of the structure is 20 μm.

Figure 5 shows a microscopic photo taken at the edge of the area irradiated with the SH of the Nd:YAG laser. Non-carbonized (red) and carbonized (black) areas can be seen.

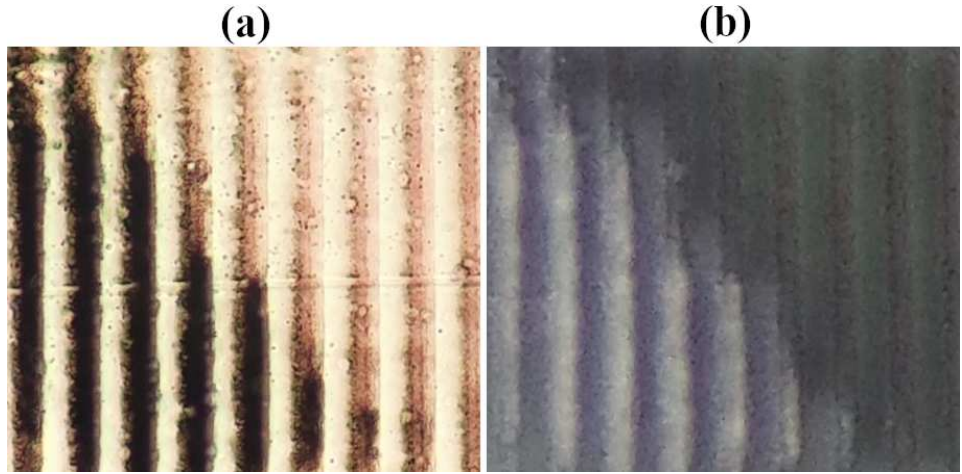

**Fig. 5** Microscopic image taken at the border of carbonized (black lines) and non-carbonized (red lines) areas (a) when illuminated with white light, (b) luminescence under UV illumination. The period of the structure is 20 μm.

It is evident that the carbonized parts of the lines are luminescent, whereas the non-carbonized pure gold containing parts are not luminescent.

## 4   Discussion

We have considered a two-step laser processing of a material in which the irradiation at the first step records an image and the irradiation at the second step leads to the development of the image resulting in the formation of a final pattern.

The specific feature here is that the developing irradiation does not affect the part of the material that has not been irradiated at the first, recording step.

The recording is conducted by short-wave light providing activation of the material, making it sensitive to the effect of developing radiation. The development is typically performed by more powerful radiation.

In the present work, the recording process is photoinduced formation of structures possessing gold nanoparticles within a polymer matrix. It employs an UV laser beam of the third harmonic of an Nd:YAG laser. The photochemical decomposition of a gold precursor takes place at this step. We have demonstrated that using the mask it is possible to record a micrograting containing alternating regions of nanocomposite material – initial material. The development process is implemented by powerful radiation of the second harmonic of the Nd:YAG laser. The wavelength of the second harmonic enters the plasmon resonance band of gold nanoparticles. This provides uniform irradiation of the initial grating, resulting in strong heating of the nanoparticles and the polymer material in its vicinity. This heating leads to carbonization. The development demonstrates blackening of the nanocomposite regions, keeping the areas of the native polymer intact. After the development, the grating looks like the black carbonized lines and lines of the virgin material. It is important that the carbonized regions are luminescent.

Thus, the development generates a pattern with strongly changed the optical properties of the initially recorded one. It is shown that at the edge of the developing laser spot it is possible to obtain patterns simultaneously possessing carbonized and non-carbonized nanocomposite parts.

Thus, it is possible to create a final pattern consisting of the developed and non-developed parts, red-black ones, like in Figs. 2 and 5.

The carbonized parts consist of core-shell gold-carbon nanoobjects separated from each other. The black lines presented in the above figures are typically non-conductive because of the absence of a percolation cluster within the modified material. The following irradiation of the pattern by powerful second harmonic radiation results in the appearance of some conductivity but the luminescence vanishes.

## 5   Conclusion

Two-step laser material processing is a promising tool for laser material modifications that allows creating a complex laser induced pattern on material surfaces.

We have considered an example of such processing which can be called recording–development. At the first step, laser radiation records an initial pattern, which is then developed by more powerful (typically longer wavelength) radiation. The properties of the initially recorded and final structures may differ drastically. This opens up an opportunity for obtaining a complex pattern consisting of the developed and the initial parts.

We have irradiated a polystyrene film containing wt. 2.5-5 % of Au precursor $(Ph_3P)Au(n\text{-}Bu)$ by the third harmonic of an Nd:YAG laser and recorded patterns containing gold nanoparticles. Their red color corresponds to the plasmon resonance within gold nanoparticles.

The homogeneous irradiation of these patterns by powerful radiation of the SH of the same laser resulted in the development of these structures, which manifests itself as darkening of the red elements. This corresponds to carbonization of the matrix near the nanoparticles. Thus, we have obtained nanocomposites with carbon nanoparticles. The carbonized elements of the pattern are luminescent.

The irradiation of part of the initially recorded structure allows obtaining a pattern consisting of both, elements with gold nanoparticles and with gold-carbon luminescent nanoobjects simultaneously.

The opportunity to create patterns consisting of elements with strongly different optical properties may be promising for photonics applications.

*Disclosures*

The authors declare that there are no financial interests, commercial affiliations, or other potential conflicts of interest that could have influenced the objectivity of this research or the writing of this paper.

*Code, Data, and Materials Availability*

Data sharing is not applicable to this article, as no new data were created or analyzed.

*Acknowledgments*

This work was supported by the Ministry of Science and Higher Education of the Russian Federation (FSWR-2020- 0035).